\documentclass[aps,prb,reprint,groupedaddress]{revtex4-1}

\usepackage{graphicx}

\begin{document}



\title{Digital Plasmonics}



\author{B. Gjonaj}

\affiliation{FOM-Institute for Atomic and Molecular Physics AMOLF,
Science Park 104, 1098 XG Amsterdam, The Netherlands}

\author{J. Aulbach}
\affiliation{FOM-Institute for Atomic and Molecular Physics AMOLF,
Science Park 104, 1098 XG Amsterdam, The Netherlands}

\author{P. M. Johnson}
\affiliation{FOM-Institute for Atomic and Molecular Physics AMOLF,
Science Park 104, 1098 XG Amsterdam, The Netherlands}

\author{A. P. Mosk}
\affiliation{Complex Photonic Systems, Faculty of Science and
Technology, and MESA+ Institute for Nanotechnology, University of
Twente, PO Box 217, 7500 AE Enschede, The Netherlands}

\author{L. Kuipers}
\affiliation{FOM-Institute for Atomic and Molecular Physics AMOLF,
Science Park 104, 1098 XG Amsterdam, The Netherlands}

\author{A. Lagendijk}
\email[]{e-mail:  a.lagendijk@amolf.nl} \affiliation{FOM-Institute
for Atomic and Molecular Physics AMOLF, Science Park 104, 1098 XG
Amsterdam, The Netherlands}
\begin{abstract}
\end{abstract}
\maketitle
\section{MANUSCRIPT}

\textbf{The field of plasmonics \cite{barnes_surface_2003} offers a
route to control light fields with metallic nanostructures through
the excitation of Surface Plasmon Polaritons (SPPs)
\cite{ozbay_plasmonics:_2006,polman_applied_2008}. These surface
waves, bound to a metal dielectric interface, tightly confine
electromagnetic energy \cite{schuller_plasmonics_2010}. Active
control over SPPs has potential for applications in sensing
\cite{prodan_hybridization_2003}, photovoltaics
\cite{atwater_plasmonics_2010}, quantum communication
\cite{altewischer_plasmon-assisted_2002,akimov_generation_2007},
nano circuitry
\cite{ebbesen_surface-plasmon_2008,engheta_circuits_2007},
metamaterials \cite{liu_plasmonic_2009,ergin_three-dimensional_2010}
and super-resolution microscopy
\cite{fang_sub-diffraction-limited_2005}. We achieve here a new
level of control of plasmonic fields using a digital spatial light
modulator. Optimizing the plasmonic phases via feedback we focus
SPPs at a freely pre-chosen point on the surface of a nanohole array
with high resolution. Digital addressing and scanning of SPPs
without mechanical motion will enable novel interdisciplinary
applications of advanced plasmonic devices in cell microscopy,
optical data storage and sensing.}

Positioning and focusing waves in transparent media requires fine
tuning of the phase profile so that waves converge and
constructively interfere at a point. A conventional lens uses
refraction to redirect the waves to the focus and a well-designed
lens shape to align the phase vectors of these waves. Due to the
fixed geometric shape of the lens, the position of the focus can
only be controlled by mechanically moving the lens or changing the
angle of incidence of the incident beam. Focusing and controlling
the position where waves constructively interfere in complex
structures require new methods that are more versatile. Optical
wavefront shaping has became a popular method that allows to focus
light even inside completely disordered materials
\cite{VellekoopOL08, cizmar_in_2010}.

Positioning and focusing SPP waves in a controlled way is important
for nanophotonic applications. To date plasmonics offers only a
limited flexibility in the control of light fields: as with the
conventional lens the geometry is typically fixed, so for a given
optical frequency the locations of optical field enhancement are
also fixed. Recently, some breakthroughs have been made on active
control in which the intensity of the light fields is influenced in
time, either through pump-probe
\cite{dionne_plasmostor:metaloxidesi_2009, macdonald_ultrafast_2009,
utikal_all-optical_2010} or through coherent control
\cite{durach_toward_2007}. Only in specific cases this control also
leads to spatial selectivity \cite{aeschlimann_adaptive_2007,
li_highly_2008, volpe_controllingoptical_2009}. However, in the
experiments the spatial selectivity is limited to a few modes
predefined by the sample structure.

We demonstrate here a new level of control of SPP wavefronts. This
control allows us to tune any SPP interference phenomenon with
unprecedented flexibility. Specifically, we show that we can
generate, focus SPPs and scan the focus on a nanohole array with an
electronically controlled spatial light modulator and standard
helium neon laser. Because the light-to-SPP conversion process is
coherent, the structured optical wavefront is projected onto the SPP
wavefront. This conversion gives us full phase control of the SPPs,
allowing us to shape the SPP wavefronts digitally. Because we use
optimization loops to determine the necessary wavefront, our method
is applicable to any plasmonic structure. Such flexible and digital
control of SPPs is a large step forward towards interdisciplinary
applications of advanced plasmonics.

The sample is a nanohole array, similar to those used typically for
Enhanced Optical Transmission (EOT) experiments
\cite{garcia-vidal_light_2010} and recently suggested for
super-resolution \cite{sentenac_subdiffraction_2008}. Our sample is
composed of a 200 nm of gold film deposited on top of 1 mm BK7 glass
substrate. The array covers an area of 30 x 30 ${\mu}\texttt{m}^{2}$
and the hole period is 450 nm. Square holes were milled with sides
of 177 nm. The SPP wavelength at the gold-air interface from
incident radiation of $\lambda_0$~=~633 nm is given by
\begin{equation}\label{equation1}\\
\lambda_{S}= \lambda_0
\rm{Re}\sqrt{\frac{\varepsilon_m+\varepsilon_d}{\varepsilon_m\varepsilon_d}},
\end{equation}
with $\varepsilon_m$ and $\varepsilon_d$ the dielectric constants of
gold and air, respectively. Using tabulated bulk values for
$\varepsilon_m$ \cite{johnson_optical_1972} we found
$\lambda_{S}$~=~600 nm.

Our aim is to digitally control the amplitude and phase of SPPs
locally on the surface of the sample. This control is achieved by
imaging a Spatial Light Modulator (SLM) onto the surface of the
sample thus mapping each unit cell (pixel) of the SLM to a
corresponding area on the sample. Amplitude and phase control of the
SLM is achieved via the 4-pixel technique \cite{vanPuttenSLM08}
where four adjacent pixels are grouped into a superpixel. We apply
to the SLM a 32 x 32 superpixel division and we independently
control amplitude and phase of each superpixel.

\begin{figure}[t!!!]\
\centering\includegraphics[width=0.45\textwidth]{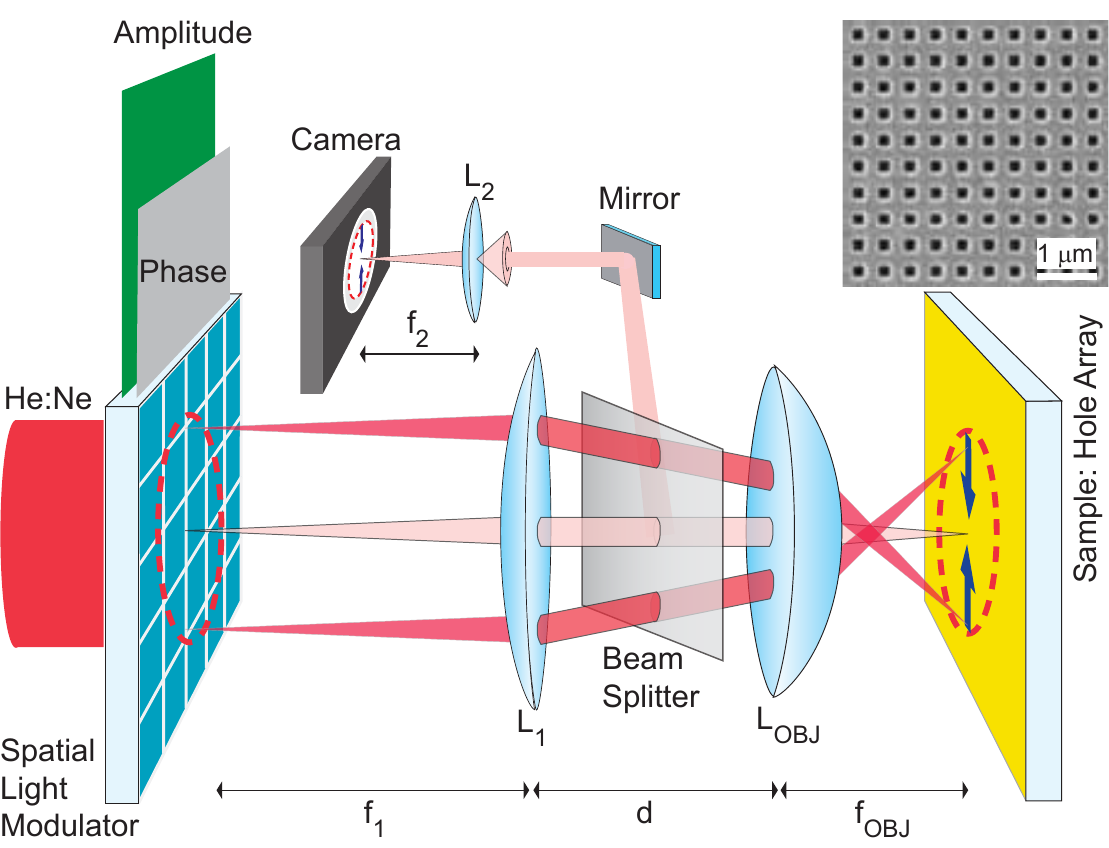}\\
\caption{Experimental Setup. The Spatial Light Modulator (SLM) is
projected onto the sample via the two-lens imaging system $L_1$ and
$L_{OBJ}$. The demagnification is 650 times. Every image point on
the sample is formed with a different average angle of incidence
(shown for three pixels). The amplitude and phase of each pixel of
the SLM are independently controlled with a computer. The sample is
a nanohole grating engraved on a gold film. The blue arrows
illustrate the propagation of Surface Plasmon Polaritons (SPP)
launched from two pixels of the SLM. The amplitudes and phases of
the SPPs are effectively clamped to those of the launching pixels.
The surface of the sample is imaged onto the camera via $L_{OBJ}$
and $L_2$.}
 \label{Figure_setup} \end{figure}

A diagram of the setup is given in Fig.~\ref{Figure_setup}. Based on
SPP momentum conservation we designed the imaging system such that
plasmons are launched toward the center. The light reflected from
the sample is imaged on the detector. This light includes both the
direct reflection of the illuminating beam and the scattered light
from SPPs. Thus the resulting image is a combination of both the SLM
amplitude pattern and the generated SPPs.

To separate plasmonic from optical effects we spatially design the
amplitude of the incident light to define four bright plasmon
launching areas and one central dark arena. Any intensity detected
inside the arena is purely plasmonic. The designed amplitude profile
for focusing experiments is a four-block pattern of fully ``on"
(A=1) superpixels on an ``off" (A=0) background.

\begin{figure}[t]
\centering\includegraphics[width=0.4\textwidth]{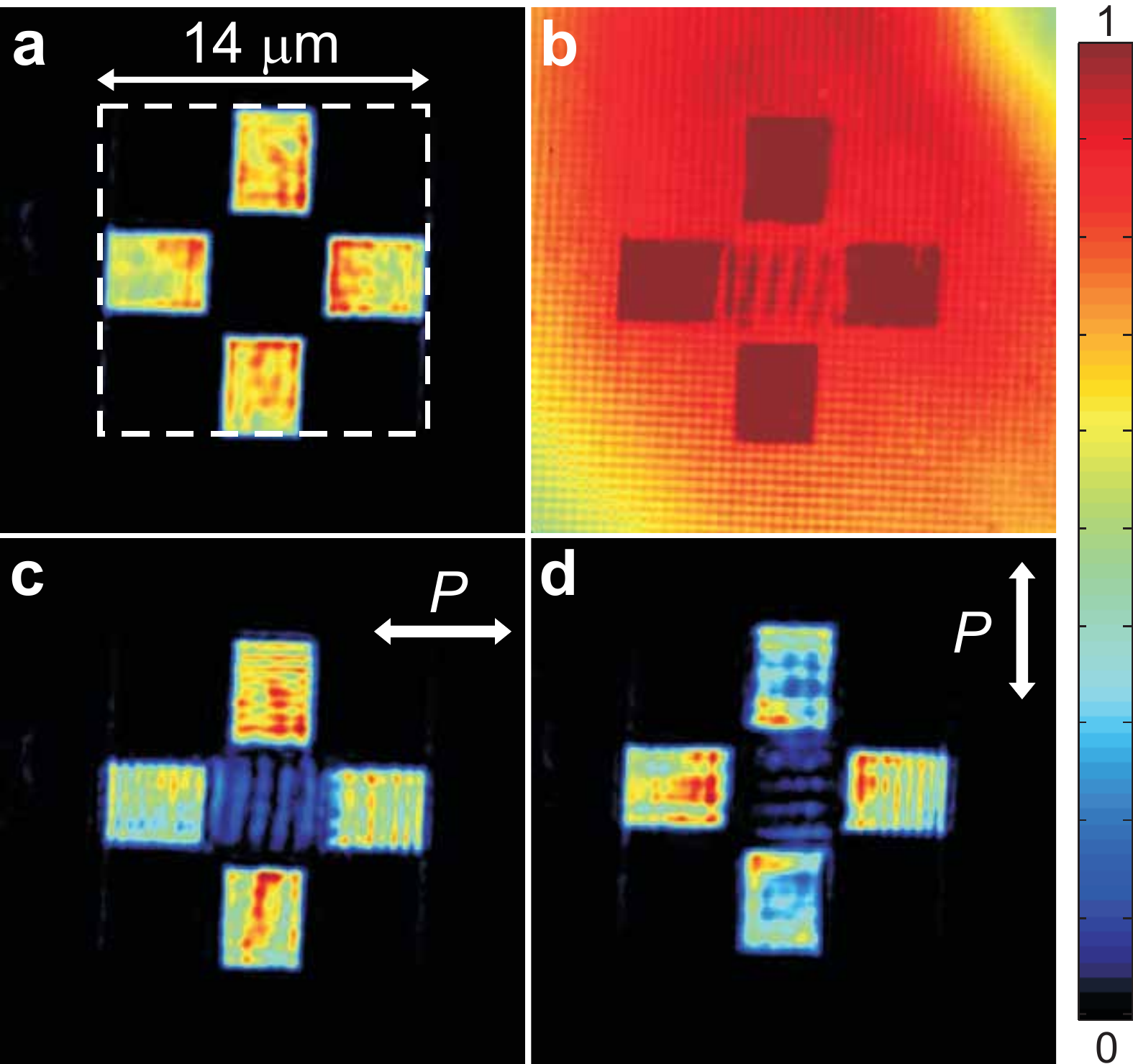}\\
\caption{Amplitude projection for uniform phase profile (no
optimization). In each image the bright rectangles are the
illuminated (amplitude=1) SPP launching areas. The SPPs are observed
in the dark (amplitude=0) central SPP arena. (\textbf{a}) Bare gold
reference (no SPPs launched). The dashed lines demarcate the SLM
area. (\textbf{b})-(\textbf{d}) SLM projected on the 450 nm hole
array. (\textbf{b}) SLM image plus white light illumination to
observe the hole array. (\textbf{c}) SPPs launched toward the
central SPP arena. (\textbf{d}) Vertical polarization of incident
light (horizontal polarization for the other images).}

\label{amplitudes}
\end{figure}

The resulting illuminated areas on a bare gold substrate are visible
in Fig.~\ref{amplitudes}a. The overall phase is constant. Each ``on"
block is 10~x~8 superpixels in size. Because no SPPs are launched on
bare gold due to momentum mismatch, the image of
Fig.~\ref{amplitudes}a is used as a background reference and measure
of contrast ratio between the ``on" and ``off" areas. The observed
contrast is nearly three orders of magnitude confirming that no
photons enter the SPP arena.

When the designed amplitude profile is projected onto the hole
array, SPPs propagate into the central dark arena. The nanohole
array has a dual role: it is used to launch SPPs (bright rectangles
in Figs.~\ref{amplitudes}b-\ref{amplitudes}d) and to visualize the
launched SPP's through their out-of-plane scattering in the central
arena. The SPPs are launched only along the direction of the
incident polarization as seen in Figs.~\ref{amplitudes}c and
\ref{amplitudes}d, consistent with expectations
\cite{van_oosten_nanohole_2010}. In Fig.~\ref{amplitudes}b, the
sample is illuminated both by the structured amplitude profile and
an additional white light source, revealing both the hole array
grating itself (the fast amplitude modulation) and the laser light.
This figure also demonstrates the optical resolution of the setup:
sufficient to resolve the presence of the hole array pattern but not
the shape of holes.

When two counterpropagating SPP waves interfere, a standing wave
pattern of intensity is created. The observed period of the fringe
pattern is clearly not half the SPP wavelength, as is expected for
SPPs propagating on an ideally smooth and non-corrugated sample. The
measured fringe period is $1\pm0.05~\mu$m. We attribute the fringe
patterns to a Moire$\acute{}$ effect between the true standing SPP
wave and the periodicity of the arrays.

\begin{widetext}

\begin{figure}[t!]\
\centering   \includegraphics[width=0.9\textwidth]{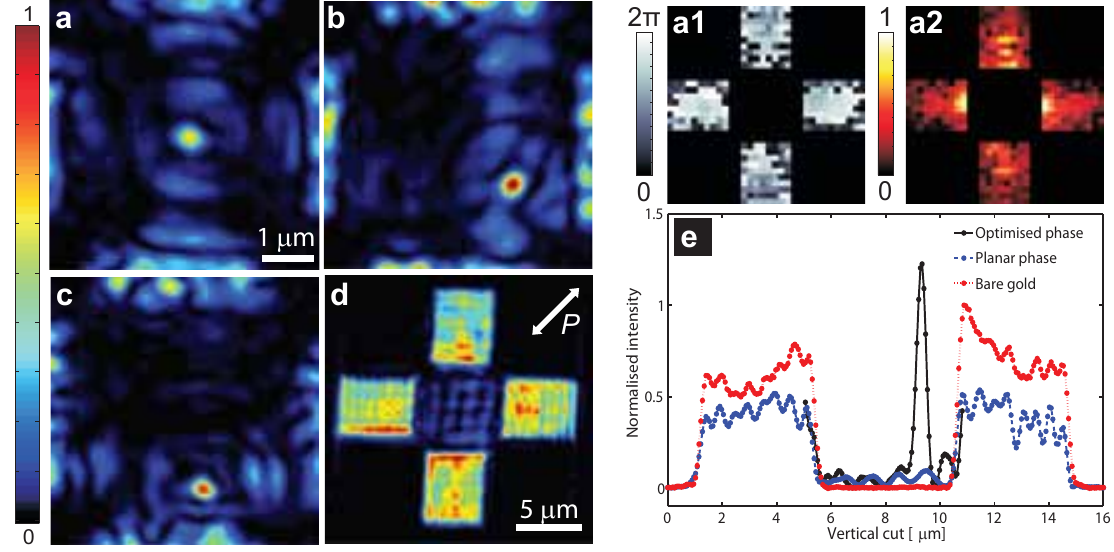}\\
\caption{Dynamic focusing of SPPs. (\textbf{a})  The relative phases
of the superpixels are optimized to focus SPPs in the center of the
SPP arena. The intensity in the target spot is purely plasmonic and
20 times higher than the average background of an unstructured
plasmonic wavefront. The focus size is diffraction limited by the
detecting optics. (\textbf{b}) and (\textbf{c}) Demonstration of SPP
focusing on freely chosen targets in the SPP arena. (\textbf{d})
Background reference of an unstructured SPP wavefront (uniform phase
profile). In achieving the focus of image (\textbf{a}) we recorded
the map of optimal phases (\textbf{a1}) and of relative
contributions (\textbf{a2}) of the superpixels, respectively. Due to
reciprocity these maps coincide with the phase and amplitude Green's
function of a SPP source at the target. The amplitude map shows the
decaying nature of the SPPs. (\textbf{e}) Quantitative analysis of
the SPP focusing showing vertical cuts of (\textbf{b}) and
(\textbf{d}). These cuts are normalized to the peak intensity of the
bare gold case, also included in the graph.}
\label{optimized}\end{figure}

\end{widetext}

Now we present experiments of SPP focusing with digital phase
control. The achieved SPP focusing is shown in Fig.~\ref{optimized}.
We use a phase optimization loop \cite{VellekoopOC2008} to focus
SPPs at a pre-chosen target. This loop yields the optimal phase
$(\widetilde{\phi})$ for each superpixel as well as the relative
contribution $(C)$ to focus.The amplitude profile is the same as for
the bare gold case with four launching areas and a central dark
arena where only SPPs can propagate. The incident polarization is
diagonal with the grating lines so as to have all available angles
($2\pi$ range) contributing to the focus, thereby maximizing the NA
and resolution.

Successful focusing at center of the SPP arena is shown in
Fig.~\ref{optimized}a. The structured SPP wavefront produces an
intensity in the designated target that is at least 20 times higher
than the average SPP background of an unstructured wavefront. The
measured size of the plasmonic focus is 420 nm, consistent with the
diffraction limit of our optics. The flexibility of the method
(scanning the focus) is demonstrated in Fig.~\ref{optimized}b and
Fig.~\ref{optimized}c which show the SPP focus relocated without
mechanical motion to controlled positions in the plasmonic arena.
Thus the plasmonic arena is our field-of-view.

We interpret SPP focusing in terms of Green's functions connecting
the electric fields at any two points. We idealize every ``on"
superpixel $n$ with a light source positioned at $\textbf{r}_{n}$
and with phase $\phi(\textbf{r}_n)$ and strength
$A(\textbf{r}_n)=1$.  The amplitude of the electric field
(normalized to the incident field) at the target $\textbf{r}_{0}$
due to these sources is
\begin{equation}\label{equation3}\\
E\left(\textbf{r}_{0},\left\{\phi(\textbf{r}_n)\right\}\right)=
\sum_{n}^{N}g\left(\textbf{r}_{0},\textbf{r}_{n}\right)\exp{\left[i\phi(\textbf{r}_{n})\right]},
\end{equation}
where $g\left(\textbf{r}_{0},\textbf{r}_{n}\right)$ is the Green's
function connecting each source to the target, and the sum runs over
all the ``on" superpixels of the amplitude profile. The target field
is maximal when all source contributions are in phase. The optimal
phase for superpixel $n$ is
$\widetilde{\phi}(\textbf{r}_n,\textbf{r}_0)=-\arg[g\left(\textbf{r}_{0},\textbf{r}_{n}\right)]$.
Imposing this phase to the superpixel yields an intensity increase
of $C(\textbf{r}_n,\textbf{r}_0)=
\left|g\left(\textbf{r}_{0},\textbf{r}_{n}\right)\right|$. We can
write
\begin{equation}\label{equation4}\\
g\left(\textbf{r}_{n},\textbf{r}_{0}\right)=
C(\textbf{r}_n,\textbf{r}_0)
\exp{\left[i\widetilde{\phi}(\textbf{r}_n,\textbf{r}_0)\right]}.
\end{equation}

Equation \ref{equation4} implies that when a focus is achieved in
the SPP arena, the recorded optimal phases and relative
contributions of the superpixels give the Green's function for a
plasmonic source located at that exact focus point. Thus the
superpixels of the SLM effectively behave as amplitude and phase
sensitive detectors. These results are valid for any Green's
function or nanostructure and can be extended to the time domain
\cite{FinkPRL} and the transfer matrix approach
\cite{popoff_measuringtransmission_2010}.

For a perfectly smooth sample with no corrugations the SPP Green's
function is simply a cylindrical wave in two dimensions (the Hankel
function $H_{0}^{(1)}(Kr)$ with $K$ the complex-valued SPP
momentum). Our digitally measured Green's function includes the
light-to-SPP coupling and therefore presents much more complexity.

With digital plasmonics we demonstrate the first ``\emph{black-box}"
with nanoscale SPP outputs and text file inputs. Specifically, we
focus SPPs on hole arrays and locally scan the focus freely over a
field-of-view (SPP arena) without any mechanical translation. In
achieving such dynamic focusing we recorded amplitude and phase
Green's functions. These digital records, which contain the full
complexity of the Green's function, are used as self-calibrated
inputs. The method can be extended to any plasmonic structure and to
the time domain. This digital plasmonic workbench is anticipated to
enable interdisciplinary applications in microscopy, optical data
storage and in bio-sensing.

We thank Elbert van Putten and Jean Cesario for stimulating and
helpful discussions. For sample fabrication we thank Hans
Zeijlermaker. This work is part of the research program of the
``Stichting voor Fundamenteel Onderzoek der Materie", which is
Financially supported by the ``Nederlandse Organisatie voor
Wetenschappelijk Onderzoek".


\begin{thebibliography}{99}

\bibitem{barnes_surface_2003}
Barnes, W.~L., Dereux, A. \& Ebbesen, T.~W.
    Surface plasmon subwavelength optics.
    \emph{Nature} \textbf{424}, 6950, 824 (2003).

\bibitem{ozbay_plasmonics:_2006}
Ozbay, E.
    Plasmonics: Merging photonics and electronics at nanoscale dimensions.
    \emph{Science} \textbf{311}, 5758, 189 (2006).

\bibitem{polman_applied_2008}
Polman, A.
    {APPLIED} {PHYSICS:} plasmonics applied.
    \emph{Science} \textbf{322}, 5903, 868 (2008).

\bibitem{schuller_plasmonics_2010}
Schuller, J.~A. et al.
    Plasmonics for extreme light concentration and manipulation.
    \emph{Nat Mater} \textbf{9}, 3, 193 (2010).

\bibitem{prodan_hybridization_2003}
Prodan, E., Radloff, C., Halas, N.~J. \& Nordlander, P.
    A hybridization model for the plasmon response of complex nanostructures.
    \emph{Science} \textbf{302}, 5644, 419 (2003).

\bibitem{atwater_plasmonics_2010}
Atwater, H.~A. \& Polman, A.
    Plasmonics for improved photovoltaic devices.
    \emph{Nat Mater} \textbf{9}, 3, 205 (2010).

\bibitem{altewischer_plasmon-assisted_2002}
Altewischer, E., van Exter, M.~P. \& Woerdman, J.~P.
    Plasmon-assisted transmission of entangled photons.
    \emph{Nature} \textbf{418}, 6895, 304 (2002).

\bibitem{akimov_generation_2007}
Akimov, A.~V. et al.
    Generation of single optical plasmons in metallic nanowires coupled to quantum dots.
    \emph{Nature} \textbf{450}, 7168, 402 (2007).

\bibitem{ebbesen_surface-plasmon_2008}
Ebbesen, T.~W., Genet, C. \& Bozhevolnyi, S.~I.
    Surface-plasmon circuitry.
    \emph{Phys Today} \textbf{61}, 5, 44 (2008).

\bibitem{engheta_circuits_2007}
Engheta, N.
    Circuits with light at nanoscales: Optical nanocircuits inspired by metamaterials.
    \emph{Science} \textbf{317}, 5845, 1698 (2007).

\bibitem{liu_plasmonic_2009}
Liu, N. et al.
    Plasmonic analogue of electromagnetically induced transparency at the drude damping limit.
    \emph{Nat Mater} \textbf{8}, 9, 758 (2009).

\bibitem{ergin_three-dimensional_2010}
Ergin, T., Stenger, N., Brenner, P., Pendry, J.~B. \& Wegener, M.
    Three-Dimensional invisibility cloak at optical wavelengths.
    \emph{Science} \textbf{328}, 5976, 337 (2010).

\bibitem{fang_sub-diffraction-limited_2005}
Fang, N., Lee, H., Sun, C. \& Zhang, X.
    Sub-Diffraction-Limited optical imaging with a silver superlens.
    \emph{Science} \textbf{308}, 5721, 534 (2005).

\bibitem{VellekoopOL08}
Vellekoop, I.~M., van Putten, E.~G., Lagendijk, A. \& Mosk, A.~P.
    Demixing light paths inside disordered metamaterials.
    \emph{Opt. Express} \textbf{16}, 1, 67 (2008).

\bibitem{cizmar_in_2010}
Cizmar, T., Mazilu, M. \& Dholakia, K.
    In situ wavefront correction and its application to micromanipulation.
    \emph{Nat Photon} \textbf{4}, 6, 388 (2010).

\bibitem{dionne_plasmostor:metaloxidesi_2009}
Dionne, J.~A., Diest, K., Sweatlock, L.~A. \& Atwater, H.~A.
    PlasMOStor: a Metal-Oxide-Si field effect plasmonic modulator.
    \emph{Nano Lett} \textbf{9}, 2, 897 (2009).

\bibitem{macdonald_ultrafast_2009}
{MacDonald}, K.~F., Samson, Z.~L., Stockman, M.~I. \& Zheludev,
N.~I.
    Ultrafast active plasmonics.
    \emph{Nat Photon} \textbf{3}, 1, 55 (2009).

\bibitem{utikal_all-optical_2010}
Utikal, T., Stockman, M.~I., Heberle, A.~P., Lippitz, M. \& Giessen,
H.
    All-Optical control of the ultrafast dynamics of a hybrid plasmonic system.
    \emph{Phys. Rev. Lett.} \textbf{104}, 11, 113903 (2010).

\bibitem{durach_toward_2007}
Durach, M., Rusina, A., Stockman, M.~I. \& Nelson, K.
    Toward full spatiotemporal control on the nanoscale.
    \emph{Nano Lett} \textbf{7}, 10, 3145 (2007).

\bibitem{aeschlimann_adaptive_2007}
Aeschlimann, M. et al.
    Adaptive subwavelength control of nano-optical fields.
    \emph{Nature} \textbf{446}, 7133, 301 (2007).

\bibitem{li_highly_2008}
Li, X. \& Stockman, M.~I.
    Highly efficient spatiotemporal coherent control in nanoplasmonics on a nanometer-femtosecond scale by time reversal.
    \emph{Phys. Rev. B} \textbf{77}, 19, 195109 (2008).

\bibitem{volpe_controllingoptical_2009}
Volpe, G., Cherukulappurath, S., Parramon, R.~J., {Molina-Terriza},
G. \& Quidant, R.
    Controlling the optical near field of nanoantennas with spatial Phase-Shaped beams.
    \emph{Nano Lett} \textbf{9}, 10, 3608 (2009).

\bibitem{garcia-vidal_light_2010}
Garcia-Vidal, F.~J., Martin-Moreno, L., Ebbesen, T.~W. \& Kuipers,
L.
    Light passing through subwavelength apertures.
    \emph{Rev. of Mod. Phys.} \textbf{82}, 1, 729 (2010).

\bibitem{sentenac_subdiffraction_2008}
Sentenac, A. \& Chaumet, P.~C.
    Subdiffraction light focusing on a grating substrate.
    \emph{Phys. Rev. Lett.} \textbf{101}, 1, 013901 (2008).

\bibitem{johnson_optical_1972}
Johnson, P.~B. \& Christy, R.~W.
    Optical constants of the noble metals.
    \emph{Phys. Rev. B} \textbf{6}, 12, 4370 (1972).

\bibitem{vanPuttenSLM08}
van Putten, E.~G., Vellekoop, I.~M. \& Mosk, A.~P.
    Spatial amplitude and phase modulation using commercial twisted nematic lcds.
    \emph{Appl. Opt.} \textbf{47}, 12, 2076 (2008).

\bibitem{van_oosten_nanohole_2010}
van Oosten, D., Spasenovic, M. \& Kuipers, L.
    Nanohole chains for directional and localized surface plasmon excitation.
    \emph{Nano Lett} \textbf{10}, 1, 286 (2010).

\bibitem{VellekoopOC2008}
Vellekoop, I. \& Mosk, A.
    Phase control algorithms for focusing light through turbid media.
    \emph{Opt. Commun.} \textbf{281}, 11, 3071 (2008).

\bibitem{FinkPRL}
Derode, A., Roux, P. \& Fink, M.
    Robust acoustic time reversal with high-order multiple scattering.
    emph{Phys. Rev. Lett.} \textbf{75}, 23, 4206 (1995).

\bibitem{popoff_measuringtransmission_2010}
Popoff, S.~M. et al.
    Measuring the transmission matrix in optics: An approach to the study and control of light propagation in disordered media.
    \emph{Phys. Rev. Lett.} \textbf{104}, 10, 100601 (2010).

\end{thebibliography}

\end{document}